\documentclass[a4paper,10pt,twoside]{cpc-hepnp}

\usepackage{multicol}
\usepackage{graphicx}
\usepackage{booktabs}
\usepackage{amssymb,bm,mathrsfs,bbm,amscd}
\usepackage[tbtags]{amsmath}
\usepackage{lastpage}

\begin{document}

\fancyhead[co]{\footnotesize LIU Long-Xiang~ et al: The study of Borromean halo nuclei by the neutron wall with simulation}

\footnotetext[0]{Received 12 January 2013}

\title{The study of Borromean halo nuclei by the neutron wall with simulation \thanks{Supported by National Natural Science Foundation of China under Gr.10775159, 10925526 and 11079044 }}

\author{%
      LIU Long-Xiang(ÁõÁúÏé)$^{1,2,3}$
\quad SUN Zhi-Yu(ËïÖ¾Óî)$^{1;1)}$\email{sunzhy@impcas.ac.cn}%
\quad YUE Ke(ÔÀçæ)$^{1}$
\quad XIAO Guo-Qing(Ф¹úÇà)$^{1}$\\
\quad CHEN Xi-Meng(³ÂÎõÃÈ)$^{2}$
\quad YU Yu-Hong(ÓàÓñºé)$^{1}$
\quad ZHANG Xue-Heng(ÕÂѧºã)$^{1}$
\quad WANG Shi-Tao(ÍõÊÀÌÕ)$^{1}$\\
\quad TANG Shu-Wen(ÌÆÊöÎÄ)$^{1,3}$
\quad ZHOU Yong(ÖÜÓÂ)$^{1,2,3}$
\quad YAN Duo(ãÆîì)$^{1,3}$
\quad FANG Fang(·½·¼)$^{1}$
}
\maketitle

\address{%

$^1$ Institute of Modern Physics, Chinese Academy of Sciences, Lanzhou 730000, China\\
$^2$ The School of Nuclear Science and Technology Lanzhou University, Lanzhou 730000, China\\
$^3$ University of Chinese Academy of Sciences, Beijing 100049, China\\
}

\begin{abstract}
The model of three-body Borromean halo nuclei breakup was described by using standard phase space distributions and the Monte Carlo simulation method was established to resolve the detection problem of two neutrons produced from breakup reaction on the neutron wall detector. For $^{6}$He case, overall resolution $\sigma_{E_{k}}$ for the Gaussian part of the detector response and the detection efficiency including solid angle acceptance with regard to the excitation energy $E_{k}$ are obtained by the simulation of two neutrons from $^{6}$He breakup into the neutron wall. The effects of the algorithm on the angular and energy correlations of the fragments are briefly discussed.
\end{abstract}



\begin{multicols}{2}

\section{Introduction}

The neutron wall~\cite{xu,yu}, which is located at the external target experimental hall of CSRm at the Institute of Modern Physics, Chinese Academy of Sciences, has been built to measure neutrons in near-relativistic heavy-ion collisions by the time of flight (TOF) method. The main aim of the neutron wall is the complete measurement of neutron momenta to reconstruct the invariant mass of the nucleus, which is feasible due to strong kinematical focussing in the case of neutron emission from excited projectiles.

 There is a specific problem which is related to the neutron detection. A number of the neutron wall's submodules are fired by neutrons impinging on it and a pattern recognition algorithm must be employed in order to disentangle multiple neutron hits. The algorithm and its performance of one neutron, under the circumstances of an simulation very similar to the present one, is described in Ref.~\cite{liu}. When two neutrons need to be detected by the neutron wall at the same time, the effects of the algorithm on the angular and energy correlations of the fragments must be studied in detail, and the solid angle acceptance of two neutrons also need to be considered. The main effect takes place in a reduced double-hit recognition capability in the case where two neutrons interact in close vicinity  to each other in the neutron wall. In three-body Borromean halo nuclei~\cite{halo} breakup experiments, it has to be corrected for such detection deficiencies on the basis of realistic event simulations, adjusted to the experiments~\cite{aum}.

One simple three-body Borromean halo nuclei breakup model was presented in the following. As an example, $^{6}$He breakup reaction will be studied to confirm the simulation method and the recognition algorithm in this paper.

\section{The model of three-body Borromean halo nuclei breakup}

Each event of three-body Borromean halo nuclei breakup is characterized by nine parameters: three components of the momentum vectors($\textbf{k}_{1},\textbf{k}_{2},\textbf{k}_{3}$) for the two neutrons and  the core-particle in the projectile rest frame~\cite{lv}. In a three-body continuum state, the fragment relative motion can be described by two relative Jacobi momenta in the following: $\hbar\mathbf{k}_{x}$ between two constituents and $\hbar\mathbf{k}_{y}$ between the third constituent and the center of mass  (c.m.) of the first pair. Then, the normalized Jacobi momentum coordinates are constructed as

\begin{eqnarray}\label{jacobi}
\mathbf{k}_{x}=\mu_{x}(\frac{\mathbf{k}_{1}}{m_{1}}-\frac{\mathbf{k}_{2}}{m_{2}}),\hphantom{000000}{\mu}_{x} =\frac{m_{1}m_{2}}{(m_{1}+m_{2})},\hphantom{000}\nonumber\\[5mm]
\mathbf{k}_{y}=\mu_{y}(\frac{\mathbf{k}_{3}}{m_{3}}-\frac{\mathbf{k}_{1}+\mathbf{k}_{2}}{m_{1}+m_{2}}),\hphantom{00}\mu_{y} =\frac{m_{3}(m_{1}+m_{2})}{M},\\[5mm]
\mathbf{k}_{c.m.}=\mathbf{k}_{1}+\mathbf{k}_{2}+\mathbf{k}_{3}.\hphantom{00000000000000000000000}\nonumber
\end{eqnarray}
where $m_{i}$ and $\hbar\mathbf{k}_{i}$, $i=$ 1, 2, 3 are masses and momentum vectors, respectively, of the particle $i$ in the projectile rest frame and $\mu_{x}$, $\mu_{y}$ are the reduced masses, $M=m_{1}+m_{2}+m_{3}$.

The sum of kinetic energies for the fragments' relative motions, $E_{k}=E_{x}+E_{y}=\hbar^{2}\mathbf{k}^{2}_{x}/2\mu_{x}+\hbar^{2}\mathbf{k}^{2}_{y}/2\mu_{y}$ is equal to the excitation energy $E_{k}$ of the three-body Borromean halo system above threshold. The energy $E_{k}$ fixes only a total phase volume which is accessible for the fragments and fragment kinetic energies have continuous distributions within this volume. Two variables need to be introduced for future reference: the variable $\varepsilon=E_{x}/E_{k}$ ($E_{y}/E_{k}=1-\varepsilon$) describes the share of the relative kinetic energy residing within a pair of particles at energy $E_{k}$, and the variable $\cos\theta_{xy}=(\hat{\mathbf{k}}_{x}\cdot\hat{\mathbf{k}}_{y})$ describes fragment angular correlations. In the c.m. coordinate system, where the total nucleus is at rest $(\mathbf{k}_{c.m.}=0)$, the momentum of the third particle $\hbar\mathbf{k}_{3}$ is equal to $\hbar\mathbf{k}_{y}$.

\begin{center}
\includegraphics[width=7cm]{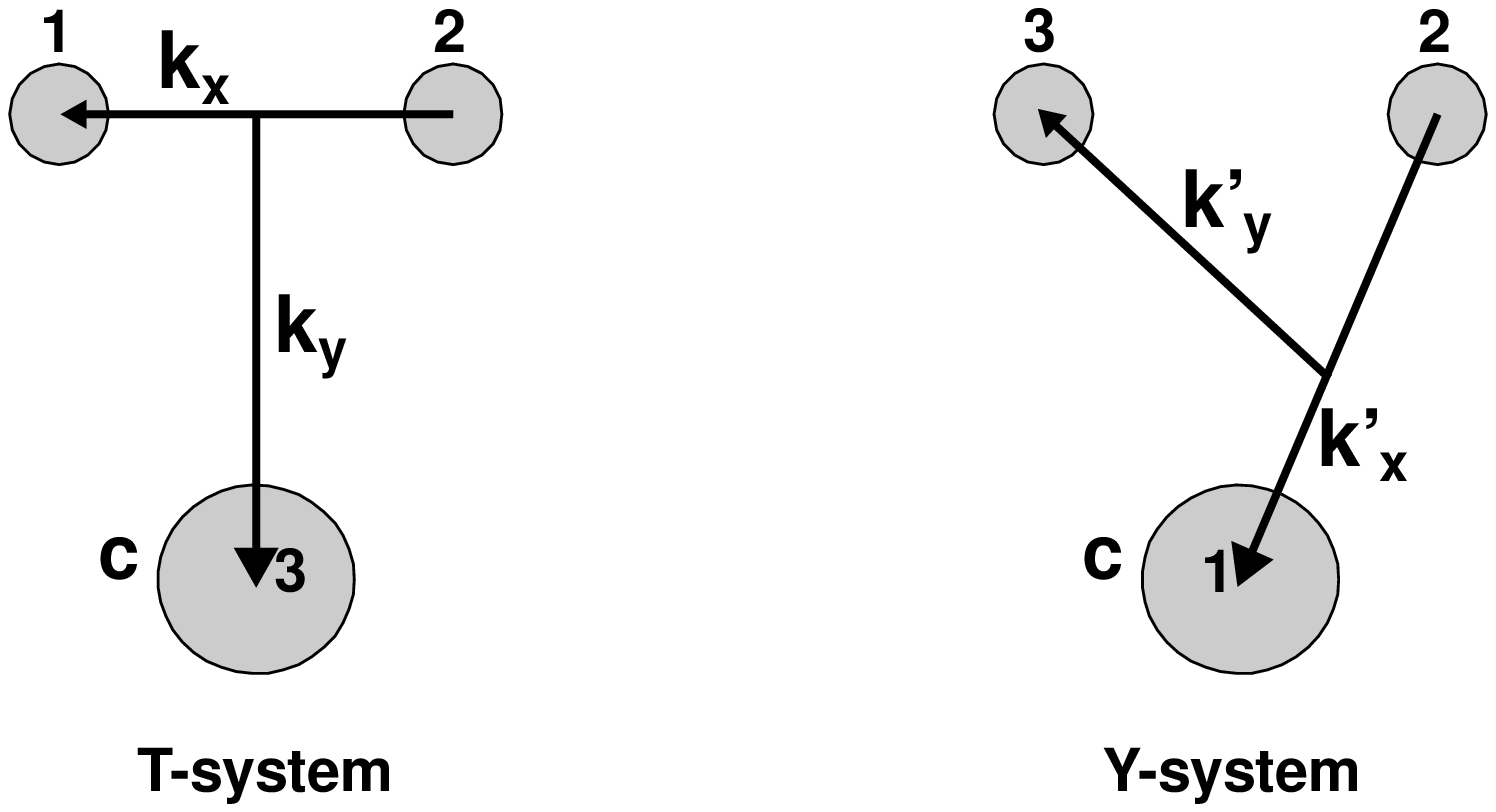}
\figcaption{\label{jaco} The \emph{\textbf{T}} and \emph{\textbf{Y}} Jacobi coordinate systems for three-body Borromean halo nuclei with two identical nucleons and the core \emph{\textbf{C}}.~\cite{lv}}
\end{center}

As is shown in Fig.~\ref{jacobi}, the case where two  of the three constituents are identical particles is discussed , implying only two different Jacobi coordinate system. The \emph{\textbf{T}} coordinate system is `cluster' representation, where the two neutrons have index 1 and 2 (coordinate $\textbf{k}_{x}$) and the core 3 is connected to the c.m. of the two neutrons, 1 and 2 (coordinate $\textbf{k}_{y}$). The \emph{\textbf{Y}} coordinate system is `shell-model' representation, where the core have index 1 and is coupled to the neutron with index 2 (coordinate $\textbf{k}'_{x}$). The second neutron 3 is then connected with the c.m. of 1 and 2 (coordinate $\textbf{k}'_{y}$).

For the event simulation, it is assumed that the available kinetic energy is distributed among the core particle and the two neutrons according to standard phase space distributions and zero momentum  transfer to the excited projectile. The fractional energy and angular correlations in the  $core+n+n$ system at fixed $E_{k}$ can be described by a probability distribution $\mathcal{W}(\varepsilon,\theta)$. It represents the probability density of finding the whole system in a final state with a configuration having fixed values of $\varepsilon$ and $\theta$ in the coordinate system \emph{\textbf{T}} and \emph{\textbf{Y}}. The probability distribution is normalized to unity.
The distributions about the kinetic energy shared in the subsystems and the polar angle between Jacobi vector $\mathbf{k}_{x}$ and $\mathbf{k}_{y}$ in the model are given by
\begin{eqnarray}\label{distribution}
  d\mathcal{W}/d\varepsilon&\sim&\sqrt{\varepsilon(1-\varepsilon)}, \nonumber\\[5mm]
  d\mathcal{W}/d\cos\theta_{xy}&\sim&constant.
\end{eqnarray}

\section{The Monte Carlo simulation method}

In order to obtain the experimental performance of the neutron wall, a Monte Carlo simulation code based on \textsc{Geant4}~\cite{g4} was implemented. $^{6}$He is one of simple three-body Borromean halo nuclei with a lot of experimental results ~\cite{lesson,aum} which can be a good reference for the simulation code. Therefore, the response of the neutron wall with selected excitation energies of $^{6}$He, disintegrating into $\alpha$+n+n, was studied by simulation. It can be prepared for the other similar three-body Borromean halo nuclei experiments on the neutron wall in the future.

The relative setup in the simulation is the same to the one described in Ref.~\cite{liu} but the distribution of the neutrons that incident into the neutron wall. The neutrons are generated from the model mentioned above without considering the simulation of the $\alpha$ particle. It can be considered that the $\alpha$ particles are completely detected with no deviation. The three-body phase space $\sqrt{E_{x}E_{y}}dE_{x}dE_{y}$ and $E_{k}$ are invariants, i.e. independent of the Jacobi system. Therefore, \emph{\textbf{T}} Jacobi coordinate system for $^{6}$He with the two neutrons and the $\alpha$ particle is selected for the simulation. The selected excitation energies of $^{6}$He breakup into $\alpha+n+n$ are 0.5, 1, 1.5, 2, 3, 4, 5 and 6 MeV according to Ref.~\cite{lesson}. The energy of $^{6}$He ion is 300 MeV/nucleon.

\section{Results and discussion }

Taking into account intrinsic detection efficiencies, position resolutions, the TOF resolution, and finite acceptances for the two neutrons, the response for a given excitation energy of $^{6}$He and subsequent decay into $\alpha+n+n$ is derived by event simulation described above. The algorithm to disentangle the two neutrons impinging on the neutron wall is identical for the analysis of simulated events in  Ref.~\cite{liu}.

\begin{center}
\includegraphics[width=7cm]{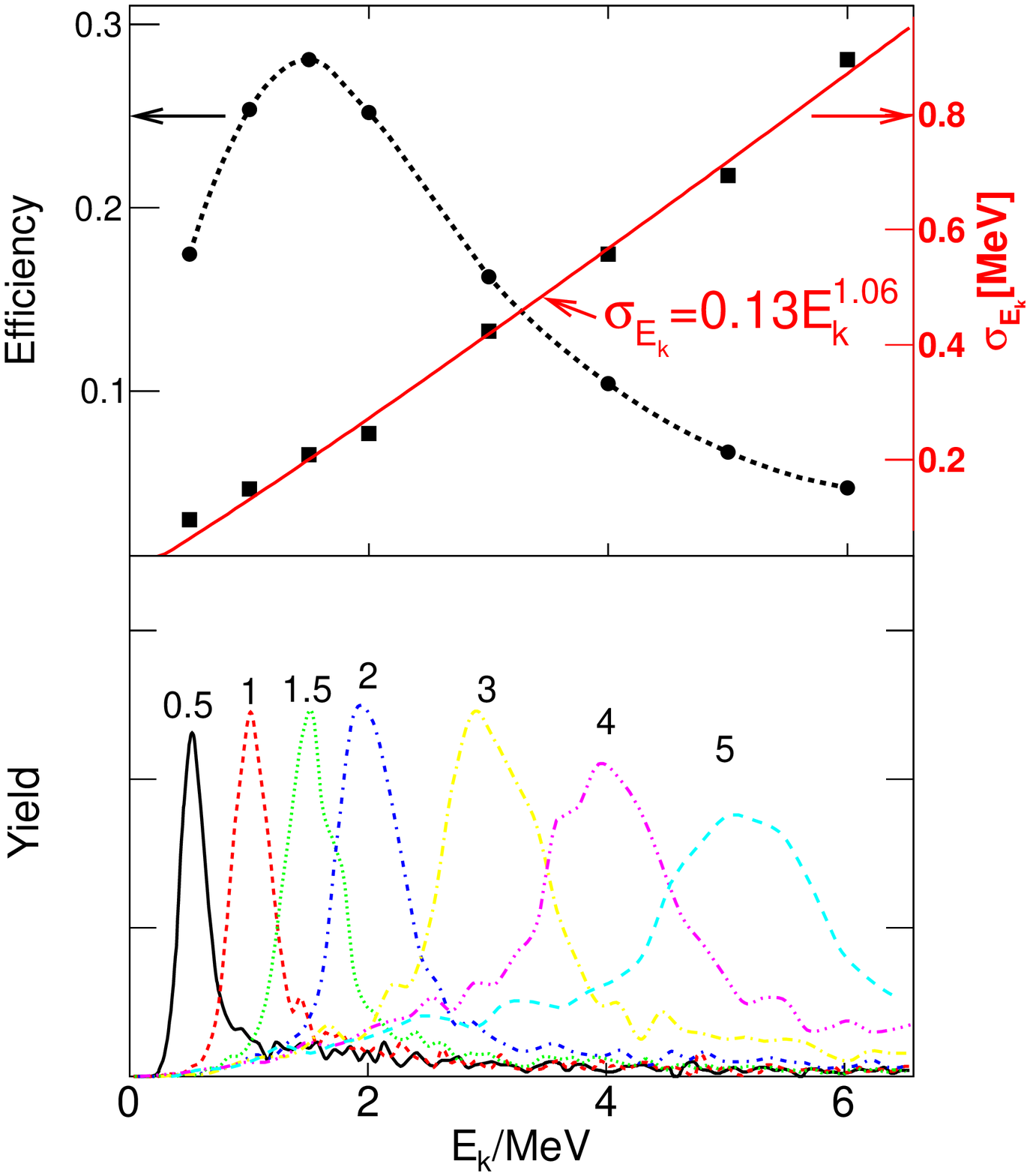}
\figcaption{\label{efficiency} (color online) (Upper panel): Resolution $\sigma_{E_{k}}$ for the Gaussian part of the neutron wall response (right hand scale, square symbols) and detection efficiency $\epsilon(E_{k})$ including solid angle acceptance (left hand scale, circle symbols). (Lower panel): Response of the neutron wall for selected excitation energies $E_{k}=$ 0.5, 1, 1.5, 2, 3, 4 and 5 MeV of $^{6}$He, disintegrating into $\alpha+n+n$ (see text). }
\end{center}

With this simulation procedure, we obtain the detection efficiencies shown in the upper panel of Fig.~\ref{efficiency}. It is observed that two neutrons impinging on the detector are well recognized in about $30\%$ of the events on the maximum. We also like to point out, that even two neutrons incident with zero relative angle can be disentangled in about $15\%$ of such cases. This can be qualitatively understood by comparing the depth of the "shower" volume (about 60 cm) with the total depth of the detector (100 cm) and taking into account the mean interaction length of about 54 cm for energetic neutrons in the neutron wall. Therefore, a differential efficiency dependent on the neutron-neutron relative angle arises which e.g. is reflected, and to most extent determines the efficiency shown in Fig.~\ref{efficiency}. The apparent decrease in efficiency, at low $E_{k}$, is due to the limited capability to resolve two neutrons with a small relative distance in the neutron wall. The decrease at higher excitation energies is due to the finite solid angle acceptance.

The same procedure also delivers response functions which can be well described by Gaussian distributions (Fig.~\ref{efficiency}, lower panel). Small non-Gaussian wings are presented on a few percent level which can be neglected under most circumstances. For the Gaussian part, we observe a variance $\sigma_{E_{k}}(MeV)\approx0.13(E_{k})^{1.06}$ with $E_{k}$ in MeV as shown in the upper panel of Fig.~\ref{efficiency}.

A comparison of the angular and energy correlation distributions for the excitation energies $E_{k}=$ 1, 2 and 5 MeV  obtained in our analysis with those from the model described above is shown in Fig.~\ref{theps}. The left column shows the angular correlation distributions depending on $\cos\theta_{xy}$ in the \emph{\textbf{T}} system; the right column gives the energy correlation ones. For the angular correlations, the difference between the detected and reconstructed event distributions after the analysis, and the model ones at the three energies is not obvious, but the most striking feature is that the reconstructed results in a concave angular distribution for $E_{k}=$ 1 MeV (Fig.~\ref{theps}(the top panel in the left column)). This is the result of the low excitation energy, which is relative to the reconstructed error.

For the  energy correlation distributions, shown in the right column, the decrease in the probability distributions between the model and the detected events at low $\varepsilon$ for $E_{k}=$ 1 and 2 MeV, is due to the limited capability to resolve two  neutrons with a small relative distance in the neutron wall. And, the decrease at high $\varepsilon$ for $E_{k}=$ 5 MeV, is due to the finite solid angle acceptance. There is almost no difference between the detected and reconstructed event distributions at the three energies. It demonstrates that the algorithm to disentangle the two neutrons impinging on the neutron wall is correct on some level.

\end{multicols}
\ruleup
\begin{center}
\includegraphics[width=13cm]{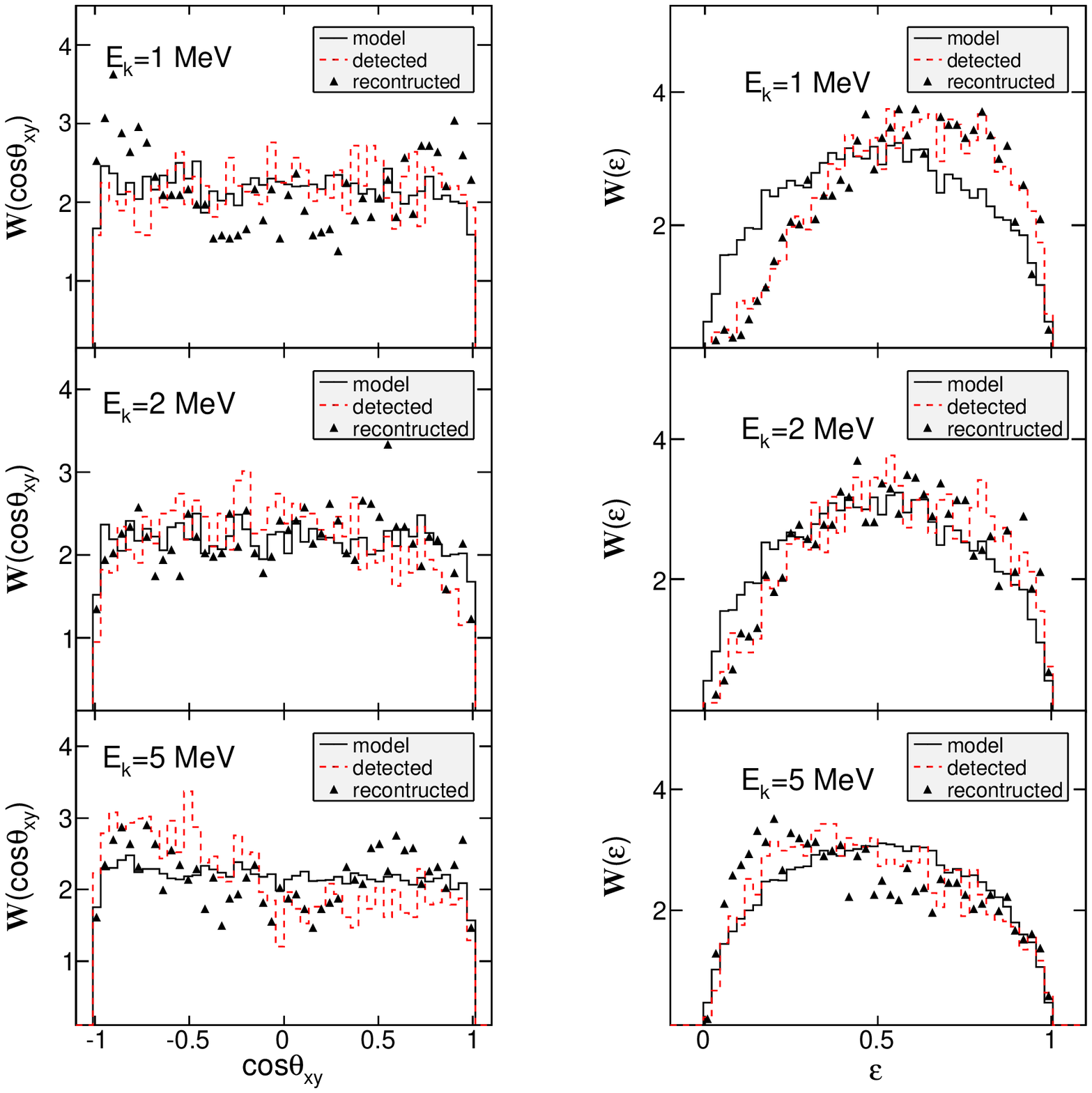}
\figcaption{\label{theps}  (color online) Projections of the probability distribution $\mathcal{W}(\varepsilon,\theta)$ in \emph{\textbf{T}} Jacobi coordinate system obtained in the selected excitation energies $E_{k}=$ 1, 2 and 5 MeV. The black solid lines correspond to the probability distributions of the standard phase model, see text; the red dashed lines correspond to the ones of the detected events; the full triangles correspond to the ones of the reconstructed events after the analysis. }
\end{center}
\ruledown

\begin{multicols}{2}

\section{Conclusion}

The model and Monte Carlo simulation method for identifying multiple neutron hits were established to resolve the detection problem of two neutrons in three-body Borromean halo nuclei breakup reaction. The three-body breakup $^{6}He\rightarrow\alpha+n+n$ was studied as an example by simulation, using 300 MeV/nuclueon $^{6}$He ion according to the model's distributions with selected excitation energies. In the simulation, the response of the neutron wall was obtained by the two neutrons based on the three-body Borromean halo nuclei breakup model. We discussed the distributions of the efficiency $\epsilon(E_{k})$ and resolution $\sigma_{E_{k}}$ with regard to the excitation energy $E_{k}$ of $^{6}$He, obtained from the event simulations. The agreement between the detected event distributions and the reconstructed ones after the analysis is reasonably good at the three energies. Therefore, the algorithm to disentangle two neutrons impinging on the neutron wall is reliable, which could be helpful for the experiments with the detection of neutrons at the external target experimental hall in the future.

\end{multicols}

\vspace{-1mm}
\vspace{2mm}

\begin{multicols}{2}

\end{multicols}

\clearpage

\end{document}